\documentclass[11pt,letterpaper]{article}
\usepackage[margin=1in]{geometry}
\usepackage{amsmath,amssymb,amsthm}
\usepackage{tikz}
\usepackage{cite}
\usepackage{titling}
\usepackage[hidelinks]{hyperref}
\predate{}\postdate{}

\newtheorem{theorem}{Theorem}
\newtheorem{lemma}{Lemma}
\newcommand{\sij}{S_{i\to j}}
\newcommand{\sumfwd}{\sum_{1\le i<j\le n}}
\newcommand{\Rplus}{\mathbb{R}_{\ge 0}}
\hypersetup{
  pdftitle={Exact Storage-Repair-Bandwidth Tradeoff for (n,k,d=k) Exact-Repair Regenerating Codes},
  pdfauthor={Chao Tian}
}
\title{Exact Storage--Repair-Bandwidth Tradeoff for 
$(n,k,d=k)$ Exact-Repair Regenerating Codes}
\author{Chao Tian}
\date{}

\begin{document}
\maketitle

\begin{abstract}
For every $n\ge2$, we establish an information-theoretic converse for
the tradeoff between storage per node and repair bandwidth in exact-repair
regenerating codes with $k=d=n-1$. The layered codes of Tian et al.,
together with space sharing, attain this bound. Thus no nonlinear code
can use less repair bandwidth at the same storage budget. The converse uses conditional copies of selected collections of node contents. Bounding what these copies jointly reveal about the file yields an induction along the segment index at fixed $n$. Applying the converse to $k+1$ nodes of an $(n,k,d=k)$ system
establishes the information-theoretic optimality of the determinant
codes of Elyasi and Mohajer for every $1\le k<n$.
\end{abstract}

\section{Introduction}\label{sec:intro}
The storage--repair-bandwidth tradeoff relates the amount of data stored
at $n$ distributed nodes to the amount downloaded (from $d$ helper nodes) to replace a failed node, where any $k$ nodes are required to allow reconstruction of the original data.
Dimakis et al.~\cite{dimakis2010network} determined the tradeoff under
functional repair, which allows the replacement node to store new data as
long as the system retains its reconstruction and repair capabilities. Exact
repair requires the replacement to reproduce the original contents of the
failed node. The characterization of the \((4,3,3)\) system in~\cite{tian4}
showed that this requirement can strictly change the tradeoff.

For systems with \(k=d=n-1\), entropy linear programs in the framework
of Yeung~\cite{yeung1997framework} gave the exact regions for \(n=4\)
and \(n=5\)~\cite{tian4,tian5}. Chen and Tian~\cite{ct2022} used a
restricted search over information inequalities to determine all but one
boundary segment for \(n=6\). A subsequent conditioning argument and exact
integer certificates established the complete regions for \(n=6\) and
\(n=7\)~\cite{tian2026finite}. In each of these cases the boundary is
attained by the layered codes of Tian et al.~\cite{layered} together with
space sharing.

For arbitrary \(n\), Prakash and Krishnan~\cite{linear} proved the
optimality of the layered tradeoff within the class of linear codes when
\(k=d=n-1\). Duursma~\cite{duursma2018shortened} obtained the linear
converse by studying shortened codes. Determinant codes give constructions
for the broader family \(k=d\le n-1\)~\cite{elyasi2016determinant}.
For codes with arbitrary encoding functions, Mohajer and Tandon~\cite{mt}
characterized the low-bandwidth portion of the tradeoff when \(k=d=n-1\)
and \(\beta\le 2\alpha/k\), where \(\alpha\) is the storage per node and
\(\beta\) is the bandwidth per helper. Sasidharan et al.~\cite{sasidharan}
obtained further information-theoretic outer bounds for general parameters.
These results leave the question of whether nonlinear codes can improve an
intermediate boundary segment of the layered tradeoff at larger \(n\).

We establish a tight converse bound for every \(n\ge 2\)
when \(k=d=n-1\). We construct an auxiliary random
vector \(Y\) by conditional sampling and bound its mutual information
with the file. Duursma's ordered repair
bound~\cite{duursma2018shortened} provided an inspiration for our approach. The resulting entropy estimates yield an induction on the boundary index \(r\) at fixed \(n\). When \(n>k+1\), we apply the converse to the \((k+1,k,k)\) system formed
by any subset of \(k+1\) nodes of an \((n,k,k)\) code. The determinant codes of
Elyasi and Mohajer~\cite{elyasi2016determinant} attain every corner of
the outer bound on the original \(n\) nodes. This extends the
tradeoff characterization for linear codes with \(k=d\) to arbitrary
exact-repair codes.

We note that the conditional sampling technique is reminiscent of the technique used in the multiple description problem \cite{ozarow1980source,tian2009approximating} and in producing non-Shannon-type inequalities \cite{zhangyeung1998,dfz2011copy}, where a new auxiliary random variable was introduced beyond what were already in the problem. Similar techniques have since found applications in other problem settings, e.g., \cite{reznic2006distortion,tian2010approximate,tian2020storage}, to yield better outer bounds in those scenarios. Though the resulting bound is exactly tight for the Gaussian multiple description problem \cite{ozarow1980source}, in other cases, the outer bounds are loose in general for the problems under consideration. Our work is the first case where the technique produces an exactly tight converse result for a lossless (zero-error, asymptotic zero-error, or network) coding problem.

\section{Notation and Achievability}\label{sec:model}
\subsection{Problem Definition and Notation}
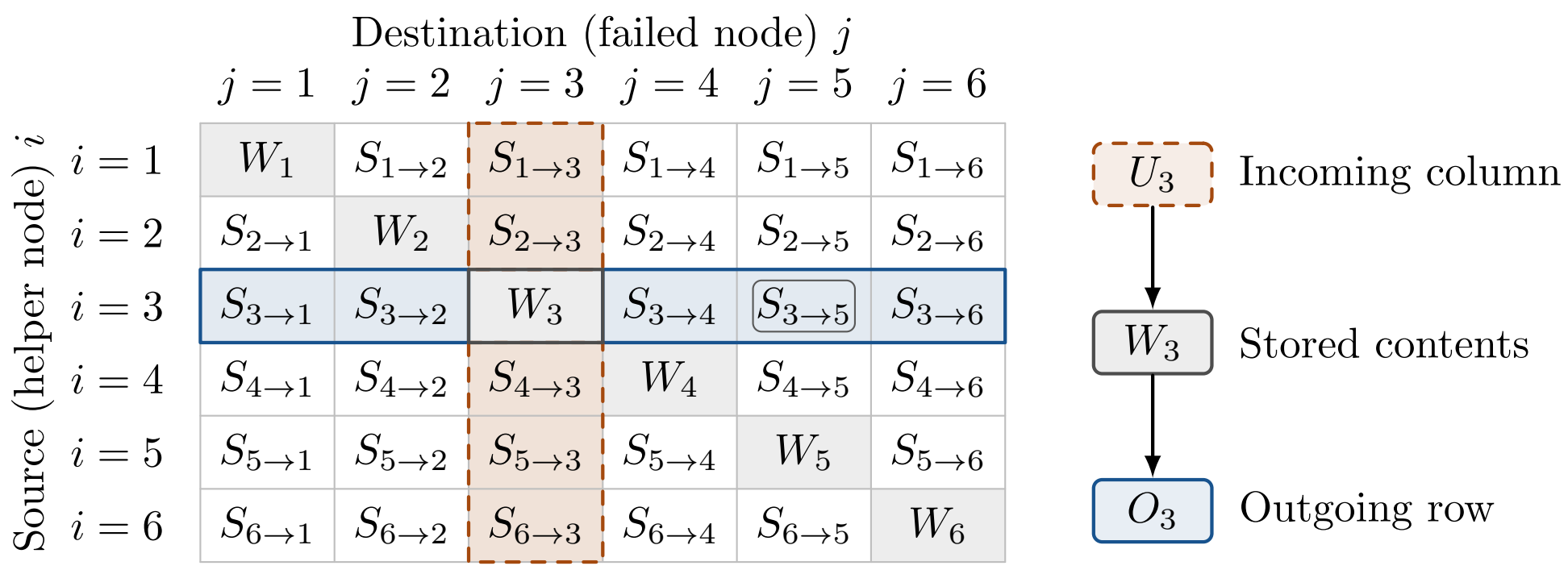
\begin{figure}[tb!]
\centering
\begingroup
\definecolor{sendblue}{RGB}{24,83,142}
\definecolor{receiveorange}{RGB}{164,73,14}
\begin{tikzpicture}[
  x=1.1cm,y=.6cm,font=\small,>=latex,line cap=round,line join=round,
  var/.style={rounded corners=2pt,minimum width=.97cm,
              minimum height=.51cm,inner sep=1pt,thick}]
  % Off-diagonal row and column of node 3. The diagonal is stored data.
  \foreach \rowstart/\rowend in {.5/2.5,3.5/6.5}{
    \fill[sendblue!12] (\rowstart,-3.5) rectangle (\rowend,-2.5);
  }
  \foreach \colstart/\colend in {-.5/-2.5,-3.5/-6.5}{
    \fill[receiveorange!16] (2.5,\colstart) rectangle (3.5,\colend);
  }
  \foreach \i in {1,...,6}{
    \fill[black!7] ({\i-.5},{-\i-.5}) rectangle ({\i+.5},{-\i+.5});
  }
  \foreach \p in {.5,1.5,...,6.5}{
    \draw[black!25] (\p,-.5)--(\p,-6.5);
    \draw[black!25] (.5,-\p)--(6.5,-\p);
  }
  \foreach \rowstart/\rowend in {.5/2.5,3.5/6.5}{
    \draw[sendblue,thick] (\rowstart,-3.5) rectangle (\rowend,-2.5);
  }
  \foreach \colstart/\colend in {-.5/-2.5,-3.5/-6.5}{
    \draw[receiveorange,thick,dashed] (2.5,\colstart) rectangle (3.5,\colend);
  }
  \draw[black!70,thick] (2.5,-3.5) rectangle (3.5,-2.5);
  \draw[black!65,rounded corners=2pt]
    (4.62,-3.35) rectangle (5.38,-2.65);
  \node at (3.5,.7) {Destination (failed node) \(j\)};
  \node[rotate=90] at (-.75,-3.5) {Source (helper node) \(i\)};
  \foreach \i in {1,...,6}{
    \node[anchor=east] at (.35,-\i) {\(i=\i\)};
    \node at (\i,0) {\(j=\i\)};
    \foreach \j in {1,...,6}{
      \ifnum\i=\j
        \node at (\j,-\i) {\(W_{\i}\)};
      \else
        \node at (\j,-\i) {\(S_{\i\to\j}\)};
      \fi
    }
  }
  % The arrows represent deterministic functions in the repair model.
  \node[var,draw=receiveorange,dashed,fill=receiveorange!10]
    (incoming) at (7.6,-1.2) {\(U_3\)};
  \node[var,draw=black!70,fill=black!7]
    (stored) at (7.6,-3.5) {\(W_3\)};
  \node[var,draw=sendblue,fill=sendblue!10]
    (outgoing) at (7.6,-5.8) {\(O_3\)};
  \draw[->,thick] (incoming.south) -- (stored.north);
  \draw[->,thick] (stored.south) -- (outgoing.north);
  \node[anchor=west] at (8.12,-1.2) {Incoming column};
  \node[anchor=west] at (8.12,-3.5) {Stored contents};
  \node[anchor=west] at (8.12,-5.8) {Outgoing row};
\end{tikzpicture}
\endgroup
\caption{Stored contents and repair messages for \(n=6\). The off-diagonal entries in column~3
form \(U_3\) and determine \(W_3\). In turn, \(W_3\) determines the
off-diagonal entries of row~3, which form \(O_3\). The outlined message
\(S_{3\to5}\) is sent by node~3 to repair node~5.}
\label{fig:repair-array}
\end{figure}

Let \([n]=\{1,\ldots,n\}\), where \(n\ge2\). The file \(F\) takes
values in a finite alphabet \(\mathcal F\). Write \(M=H(F)\) in bits,
so \(M=\log_2|\mathcal F|\) when \(F\) is uniform on \(\mathcal F\).
Node \(i\) stores \(W_i=f_i(F)\). Write \(\bar{W}_J=(W_i)_{i\in J}\)
in increasing index order and \(\bar{W}=(W_1,\ldots,W_n)\). Every
set \(J\subseteq[n]\) of \(n-1\) nodes recovers the file, so
\begin{equation}\label{eq:reconstruct}
 H(F\mid \bar{W}_J)=0 .
\end{equation}
When node \(j\) fails, helper \(i\ne j\) sends
\(\sij=\phi_{ij}(W_i)\), where $\phi_{ij}$ is a deterministic function. Node $j$ receives the tuple 
\(U_j=(S_{i\to j})_{i\ne j}\). Write \(O_i=(S_{i\to j})_{j\ne i}\) for
the outgoing tuple from helper \(i\). Exact repair means that each node
has a decoder \(\psi_j\) such that \(W_j=\psi_j(U_j)\). Thus
\begin{equation}\label{eq:repair}
 H(\sij\mid W_i)=0,\qquad H(W_j\mid U_j)=0 .
\end{equation}
All encoding and repair functions may be nonlinear. Since \(\bar{W}\)
is a function of \(F\) and recovers \(F\), we have \(H(\bar{W})=M\).
Figure~\ref{fig:repair-array} shows the stored contents and repair
messages for six nodes.

Under storage and repair budgets \(\alpha,\beta\), we have \(H(W_i)\le\alpha\) and
\(H(\sij)\le\beta\). For \(M>0\) define the normalized budgets
\(\bar{\alpha}=\alpha/M\) and \(\bar{\beta}=\beta/M\). The rate region
\(\mathcal R_n\) is the closure of the pairs
\((\bar{\alpha},\bar{\beta})\) over all uniform files, exact-repair
codes, and feasible storage and repair budgets.

\subsection{The Layered-Code Inner Bound}\label{subsec:layered}
Fix a layer size \(2\le\ell\le n\). For each \(\ell\)-subset of nodes,
the layered construction of Tian et al.~\cite{layered} takes \(\ell-1\)
independent file bits, appends their parity, and stores one symbol at
each node of the subset. Different subsets use independent bits. Any
\(n-1\) nodes recover the file because at most one symbol is missing
from each subset. To repair node \(j\), helper \(i\) sends its symbol
from every subset containing \(i\) and \(j\). The \(\ell-1\) surviving
symbols in each subset containing \(j\) recover its missing symbol.
Counting subsets gives
\[
 M_\ell=(\ell-1)\binom n\ell,\qquad
 \alpha_\ell=\binom{n-1}{\ell-1},\qquad
 \beta_\ell=\binom{n-2}{\ell-2}.
\]
The normalized storage and bandwidth are therefore
\begin{equation}\label{eq:corners}
 (\bar{\alpha}_\ell,\bar{\beta}_\ell)=\left(\frac{\alpha_\ell}{M_\ell},
                        \frac{\beta_\ell}{M_\ell}\right)
 =\left(\frac{\ell}{n(\ell-1)},\,\frac{\ell}{n(n-1)}\right),
 \qquad 2\le\ell\le n .
\end{equation}
Space sharing weights these points by the fraction of file bits assigned
to each code. Independent copies give a dense set of weights, and either
budget may be increased. Thus the layered codes achieve the region
\begin{equation}\label{eq:layered-inner}
 \mathcal L_n=\operatorname{conv}\{(\bar{\alpha}_\ell,\bar{\beta}_\ell):2\le\ell\le n\}
                  +\Rplus^2 \ \subseteq\ \mathcal R_n .
\end{equation}

For comparison with the converse, a pair \((\bar{\alpha},\bar{\beta})\in\Rplus^2\) belongs to
\(\mathcal L_n\) precisely when
\begin{align}
 (n-1)\bar{\alpha} &\ge 1,                                             \label{eq:storage}\\
 n\binom r2 \bar{\alpha}+\binom n2 \bar{\beta} &\ge \binom{r+1}{2},
                     \qquad r\in\{1,\ldots,n-1\}.           \label{eq:region}
\end{align}
Indeed, \(\bar{\alpha}_\ell\) decreases to \(\bar{\alpha}_n=1/(n-1)\), and at each layer
point the slack in~\eqref{eq:region} is
\begin{equation}\label{eq:slack}
 n\binom r2 \bar{\alpha}_\ell+\binom n2 \bar{\beta}_\ell-\binom{r+1}{2}
 =\frac{(\ell-r)(\ell-r-1)}{2(\ell-1)} .
\end{equation}
This is nonnegative for integer \(\ell,r\). For \(2\le r\le n-1\), it
vanishes at \(\ell=r,r+1\). Hence for \(\bar{\alpha}_{r+1}\le \bar{\alpha}\le \bar{\alpha}_r\), space
sharing between these two codes attains the smallest \(\bar{\beta}\) allowed by
the \(r\)-th inequality. At \(\bar{\alpha}\ge \bar{\alpha}_2\), layer size \(2\) attains
\(\bar{\beta}_2=1/\binom n2\), the minimum allowed by \(r=1\). This covers every
\(\bar{\alpha}\ge \bar{\alpha}_n\), including \(n=2\), where the sole layer point is \((1,1)\).

\section{Main Theorem}\label{sec:main}
\begin{theorem}\label{thm:region}
For every \(n\ge2\), an \((n,n-1,n-1)\) exact-repair code with file
entropy \(M\), storage \(\alpha\), and per-helper bandwidth \(\beta\)
satisfies
\[
 M\le (n-1)\alpha,\qquad
 \binom{r+1}{2}M\le n\binom r2\alpha+\binom n2\beta,\quad
 r\in\{1,\ldots,n-1\}.
\]
Consequently \(\mathcal R_n=\mathcal L_n\).
\end{theorem}

For \(M>0\), dividing the converse inequalities by \(M\) gives exactly
\eqref{eq:storage} and \eqref{eq:region}, the description of the
layered-code region in Section~\ref{subsec:layered}. For example, when
\(n=6\) the case \(r=3\) gives \(6\alpha+5\beta\ge 2M\), the segment
that completed the \((6,5,5)\) region in~\cite{tian2026finite}.

\paragraph{Remark (the \((n,k,k)\) tradeoff).}
Let \(1\le k<n\) and fix a set \(T\) of \(k+1\) nodes in an
\((n,k,k)\) code. Any \(k\) nodes in \(T\) recover the file. For each
\(j\in T\), repair \(j\) using the \(k\) helpers in \(T\setminus\{j\}\).
This choice fixes the repair maps for \(j\) even when they depend on the
helper set. The restriction to \(T\) is therefore a \((k+1,k,k)\) code
with the same file and budgets \(\alpha,\beta\). Theorem~\ref{thm:region}
gives
\[
 M\le k\alpha,\qquad
 \binom{r+1}{2}M\le (k+1)\binom{r}{2}\alpha+\binom{k+1}{2}\beta,
 \qquad 1\le r\le k .
\]
For \(m=1,\ldots,k\), the mode-\(m\) determinant code of Elyasi and
Mohajer~\cite{elyasi2016determinant} on \(n\) nodes achieves the normalized operating point
\[
 (\bar{\alpha},\bar{\beta})
 =\left(\frac{m+1}{m(k+1)},\,\frac{m+1}{k(k+1)}\right).
\]
These are precisely the corner points in~\eqref{eq:corners} with \(n\)
replaced by \(k+1\) and \(\ell=m+1\). Space sharing between adjacent
modes attains the intervening segments, so the normalized rate region
of the \((n,k,k)\) system is \(\mathcal L_{k+1}\).

\section{Proof of Theorem~\ref{thm:region}}\label{sec:proof}
We prove the converse without using the fact that the file is uniformly distributed. The induction hypothesis can then be applied to each posterior file distribution that arises under conditioning.

\subsection{Ordered Repair Under Conditioning}\label{subsec:ordered}
Fix the node order \(1,\ldots,n\). Write
\(\bar{W}_i^+=(W_{i+1},\ldots,W_n)\) for the contents of the nodes following 
sender \(i\). For receiver \(j\), let \(U_j^{-}=(S_{a\to j})_{a<j}\)
collect its incoming messages from earlier helpers. For \(1\le i<j\le n\),
write \(U_j^{i-}=(S_{a\to j})_{i<a<j}\) for the subtuple from helpers
strictly between \(i\) and \(j\). The tuples \(\bar{W}_{n}^+\) and
\(U_1^{-}\) are empty, as is \(U_j^{i-}\) when \(i=j-1\).
In Figure~\ref{fig:repair-array}, \(U_3^{-}=(S_{1\to3},S_{2\to3})\)
consists of the entries above \(W_3\) in column~3, and
\(\bar{W}_{3}^+=(W_4,W_5,W_6)\) consists of the later diagonal entries.

Let \(A\) be the sum of the node storage entropies, \(B\) the sum of
the individual repair-message entropies over all ordered pairs of
distinct nodes, and \(B^{\mathrm{u}}\) the contribution of the strict
upper triangle. Thus
\begin{equation}\label{eq:aggregate-costs}
 A=\sum_{i=1}^n H(W_i),\qquad
 B=\sum_{i\ne j}H(\sij),\qquad
 B^{\mathrm{u}}=\sumfwd H(\sij).
\end{equation}
For every positive
integer \(r \in \{1,\ldots,n-1\}\) we will show that
\begin{equation}\label{eq:entropy}
 B^{\mathrm{u}}+\binom r2 A\ \ge\ \binom{r+1}{2}M .
\end{equation}
We first prove the following instrumental lemma. For a finite random variable \(Z\) jointly distributed with the code
variables, define \(M_Z=H(F\mid Z)=H(\bar{W}\mid Z)\). A subscript
\(Z\) on \(A,B,B^{\mathrm{u}}\) conditions each summand
in~\eqref{eq:aggregate-costs} on \(Z\). In particular,
\(B_Z^{\mathrm{u}}=\sum_{i<j}H(\sij\mid Z)\).

\begin{lemma}[Conditional ordered repair bound]\label{lem:ordered}
For every such \(Z\),
\begin{equation}\label{eq:ordered}
 M_Z+\sumfwd H(\sij\mid \bar{W}_i^+,Z)\ \le\ B_Z^{\mathrm{u}} .
\end{equation}
\end{lemma}
\begin{proof}
Fix a receiver \(j\). The later nodes \(\bar{W}_j^+\) determine all
messages into \(j\) from helpers with index greater than \(j\). Thus
\((\bar{W}_j^+,U_j^{-})\) determines \(U_j\) and hence \(W_j\).
Expand the mutual information in descending order of sender index to obtain
\begin{equation}\label{eq:chain}
\begin{aligned}
 H(W_j\mid \bar{W}_j^+,Z)
 &= I(W_j;U_j^{-}\mid \bar{W}_j^+,Z)\\
 &= \sum_{i<j} I(W_j;\sij\mid \bar{W}_j^+,U_j^{i-},Z).
\end{aligned}
\end{equation}
For \(j=1\) the sum is empty and its left-hand side is zero by repair.
For \(i<j\), the tuple \(\bar{W}_i^+\) contains \(W_j\), \(\bar{W}_j^+\),
and each \(W_a\) with \(i<a<j\). Each such \(W_a\) determines
\(S_{a\to j}\), so \(\bar{W}_i^+\) also determines \(U_j^{i-}\). Consequently,
\begin{align*}
 I(W_j;\sij\mid \bar{W}_j^+,U_j^{i-},Z)
 &= H(\sij\mid \bar{W}_j^+,U_j^{i-},Z)
    -H(\sij\mid \bar{W}_j^+,U_j^{i-},W_j,Z)\\
 &\le H(\sij\mid Z)-H(\sij\mid \bar{W}_i^+,Z).
\end{align*}
Sum~\eqref{eq:chain} over \(j\) and use the reverse chain rule
\(\sum_j H(W_j\mid \bar{W}_j^+,Z)=H(\bar{W}\mid Z)\). This proves~\eqref{eq:ordered}.
\end{proof}

For $d=n-1$ and \(\ell=n\), Duursma's Theorem~1~\cite{duursma2018shortened}
reduces to
\[
 M+\sumfwd H(\sij\mid\bar{W}_i^+)
 \ \le\ \sum_{j=1}^n H(U_j^{-}\mid\bar{W}_j^+).
\]
The sum on the right is at most \(B^{\mathrm{u}}\) by subadditivity
and the fact that conditioning reduces entropy. Lemma~\ref{lem:ordered}
is a conditional version of Duursma’s ordered repair bound with a relaxed right-hand side. 
To obtain an eventual bound, 
Duursma's iteration uses a function \(V\) of the file satisfying
\(H(\sij\mid V)\le H(\sij\mid \bar{W}_i^+)\) for all \(i<j\). For linear codes he
constructs \(V\) from intersections of the sender subspaces with the spans
of the later nodes to yield an eventual characterization. 
In contrast, Section~\ref{subsec:copies} uses
conditional sampling to construct an auxiliary random vector \(Y\) that
satisfies the same inequality for every \(i<j\) and reveals at most
\(A-M\) bits about the file.

\subsection{Conditional Copies}\label{subsec:copies}
The terms \(H(\sij\mid \bar{W}_i^+)\) in Lemma~\ref{lem:ordered} have a common
conditioning tuple for messages sent by \(i\), but that tuple varies
with \(i\). Pooling these tuples would reveal \(\bar{W}_1^+\), which
already recovers \(F\) by~\eqref{eq:reconstruct}. To retain each sender's joint
distribution with its tail, we use the conditional-copy construction for entropy
inequalities~\cite{zhangyeung1998,dfz2011copy}.

\begin{lemma}\label{lem:copy}
There is a finite random tuple \(Y\) jointly distributed with the code
variables such that
\begin{equation}\label{eq:step}
 A_Y\le M,\qquad M_Y\ge 2M-A,\qquad B_Y^{\mathrm{u}}\le B^{\mathrm{u}}-M .
\end{equation}
\end{lemma}
\begin{proof}
For each \(i\), give \(Y_i\) the alphabet of \(\bar{W}_i^+\), and let
\(Y=(Y_1,\ldots,Y_n)\). Given \(\bar{W}=\bar{w}=(w_1,\ldots,w_n)\),
extend the original law by drawing \(Y\) according to the kernel
\begin{equation}\label{eq:kernel}
 p_{Y\mid \bar{W}}(y\mid \bar{w})=
 \prod_{i=1}^n p_{\bar{W}_i^+\mid W_i}(y_i\mid w_i).
\end{equation}
The kernel is needed only for \(\bar{w}\) in the support of \(\bar{W}\).
It preserves the original distribution of \(F,\bar{W}\), and the repair messages.
Given \(\bar{W}\), the
\(Y_i\)'s are independent, and the conditional distribution of \(Y_i\) depends on
\(\bar{W}\) only through \(W_i\). Marginalizing~\eqref{eq:kernel} gives
\[
 \Pr\{W_i=w_i,Y_i=y_i\}
 =\Pr\{W_i=w_i\}\Pr\{\bar{W}_i^+=y_i\mid W_i=w_i\}
 =\Pr\{W_i=w_i,\bar{W}_i^+=y_i\}.
\]
Thus \((W_i,Y_i)\) and \((W_i,\bar{W}_i^+)\) have the same distribution. 

Conditioning on all of \(Y\) and then using the reverse chain rule yields
\begin{equation}\label{eq:Aprime}
 A_Y=\sum_i H(W_i\mid Y)
 \le\sum_i H(W_i\mid Y_i)=\sum_i H(W_i\mid \bar{W}_i^+)=M .
\end{equation}
To bound the file entropy that remains, note that the product
in~\eqref{eq:kernel} gives the exact identity
\(H(Y\mid \bar{W})=\sum_i H(Y_i\mid W_i)\). Subadditivity then gives
\begin{align}
 I(\bar{W};Y)&=H(Y)-H(Y\mid \bar{W}) \notag\\
 &\le\sum_i\bigl[H(Y_i)-H(Y_i\mid W_i)\bigr]
   =\sum_i I(W_i;\bar{W}_i^+) \notag\\
 &=\sum_i H(W_i)-\sum_i H(W_i\mid \bar{W}_i^+)=A-M .               \label{eq:cost}
\end{align}
Since \(\bar{W}\) and \(F\) determine each other, \(I(\bar{W};Y)=I(F;Y)\).
Thus \(M_Y=M-I(\bar{W};Y)\ge 2M-A\). 

Since \(\sij=\phi_{ij}(W_i)\), the pairs \((\sij,Y_i)\) and
\((\sij,\bar{W}_i^+)\) have the same distribution for each \(i<j\). Hence
\[
 B_Y^{\mathrm{u}}
 \le\sumfwd H(\sij\mid Y_i)
 =\sumfwd H(\sij\mid \bar{W}_i^+)
 \le B^{\mathrm{u}}-M,
\]
where the last inequality follows from Lemma~\ref{lem:ordered} with constant \(Z\).
\end{proof}

\subsection{Proof by Induction}
We next prove inequality~\eqref{eq:entropy} by induction. Fix \(n\). At \(r=1\), inequality~\eqref{eq:entropy} is \(B^{\mathrm{u}}\ge M\),
which follows from Lemma~\ref{lem:ordered} without the conditioning $Z$ and by the nonnegativity of conditional
entropy. Let us consider the induction to $r=2$ as an illustration. Lemma~\ref{lem:ordered} gives
\(B_Y^{\mathrm{u}}\ge M_Y\) under conditioning, so~\eqref{eq:step} implies
\(B^{\mathrm{u}}-M\ge 2M-A\), or \(B^{\mathrm{u}}+A\ge3M\). This is exactly the \(r=2\) inequality.

Suppose now that~\eqref{eq:entropy} holds at a specific \(r\) for every distribution with the
stated repair and reconstruction functions. Construct \(Y\) by
Lemma~\ref{lem:copy}. For any \(y\) of positive probability, conditioning
on \(Y=y\) reweights the original file probabilities. The encoding,
reconstruction, and repair identities continue to hold on the posterior
support, since the same functions act on the same file symbols. Apply
the induction hypothesis to each such law,
multiply by \(\Pr\{Y=y\}\), and sum over \(y\). By the definition of
conditional entropy this gives
\begin{equation}\label{eq:average}
 B_Y^{\mathrm{u}}+\binom r2 A_Y\ \ge\ \binom{r+1}{2} M_Y .
\end{equation}
Using all three estimates in~\eqref{eq:step}, we obtain
\begin{equation}\label{eq:induct}
 B^{\mathrm{u}}-M+\binom r2 M
 \ \ge\ B_Y^{\mathrm{u}}+\binom r2 A_Y
 \ \ge\ \binom{r+1}{2} M_Y
 \ \ge\ \binom{r+1}{2}(2M-A).
\end{equation}
Moving the \(A\) term to the left and using
\(1+2\binom{r+1}{2}-\binom r2=\binom{r+2}{2}\) gives
\[
 B^{\mathrm{u}}+\binom{r+1}{2}A\ \ge\ \binom{r+2}{2}M .
\]
This is~\eqref{eq:entropy} at \(r+1\). The induction is therefore complete, establishing \eqref{eq:entropy}  for all \(r\in \{1,2,\ldots,n-1\}\).

The node order was arbitrary. Reversing it replaces \(B^{\mathrm{u}}\)
in~\eqref{eq:entropy} by \(B-B^{\mathrm{u}}\). Adding the inequalities
for the two orders gives
\[
 B+r(r-1)A\ \ge\ r(r+1)M .
\]
Substituting \(A\le n\alpha\) and \(B\le n(n-1)\beta\) and dividing
by \(2M\) for \(M>0\) gives~\eqref{eq:region}. Any \(n-1\) nodes recover \(F\),
so \(M\le(n-1)\alpha\), giving~\eqref{eq:storage}. All these inequalities
pass to the closure of the normalized budget pairs, so
\(\mathcal R_n\subseteq\mathcal L_n\). The opposite inclusion
in~\eqref{eq:layered-inner} completes the proof of
Theorem~\ref{thm:region}.

\section{Conclusion}\label{sec:conc}
The tradeoff implied by the layered exact repair code is optimal for
every \(n\) under arbitrary deterministic encoding and repair when $k=d=n-1$. 
The proof is extremely concise and relies on an auxiliary random variable that is outside the original problem setup. 
The same converse holds when $n-1>k=d$, establishing the information-theoretic optimality of the code by Elyasi and Mohajer. This result appears to be the first known case where the approach of introducing an expanded probability space leads to a tight characterization on an engineering problem. 

It should be noted that, for \(n=4,5,6,7\), the proof certificates use entropy inequalities on the original repair variables~\cite{tian4,tian5,tian2026finite}. It remains of interest to determine whether the converse bound can be established by Shannon-type inequalities alone. 

\section*{AI Assistance Disclosure}\label{sec:ai}

The author posed the general-\(n\) converse question using \cite{tian2026finite} as the prompt to GPT-6.2 Astra, which provided the key argument and induction step. The author suggested looking for a conditioning proof that would avoid a limiting argument in a preliminary (and more involved) version of the proof. In the ensuing interaction and collaboration, GPT-6.2 Astra identified the conditional resampling of the later nodes, derived the three estimates
in Lemma~\ref{lem:copy} and the induction on \(r\), and assisted in drafting
the manuscript. The author verified all proof steps, reorganized and revised the draft for clarity, and made the connection to related results in the literature, partly via interactions with GPT-6.2. The author takes full responsibility for the contents of the manuscript.\par

\bibliographystyle{ieeetr}
\bibliography{finite_region_references}

\end{document}